\documentclass[pdflatex,sn-mathphys]{sn-jnl}
\usepackage{tabularx}
\usepackage{setspace}
\jyear{2023}

\theoremstyle{thmstyleone}%
\theoremstyle{thmstyletwo}%

\theoremstyle{thmstylethree}%

\begin{document}

\title[]{Operability-economics trade-offs in adsorption-based CO$_2$ capture processes}


\author[1,2]{\fnm{Steven} \sur{Sachio}}\email{steven.sachio19@imperial.ac.uk}
\equalcont{These authors contributed equally to this work.}

\author[1,2]{\fnm{Adam} \sur{Ward}}\email{adam.ward16@imperial.ac.uk}
\equalcont{These authors contributed equally to this work.}

\author*[1]{\fnm{Ronny} \sur{Pini}}\email{r.pini@imperial.ac.uk}

\author*[1,2]{\fnm{Maria} M. \sur{Papathanasiou}}\email{maria.papathanasiou11@imperial.ac.uk}

\affil[1]{\orgdiv{Department of Chemical Engineering}, \orgname{Imperial College London}, \country{UK}}

\affil[2]{\orgdiv{Sargent Centre for Process Systems Engineering}, \orgname{Imperial College London}, \country{UK}}

\abstract{Low-carbon dispatchable power underpins a sustainable energy system, providing load balancing complementing wide-scale deployment of intermittent renewable power. In this new context, fossil fuel-fired power plants must be coupled with a post-combustion carbon capture (PCC) process capable of highly transient operation. To tackle design and operational challenges simultaneously, we have developed a computational framework that integrates process design with techno-economic assessment. The backbone of this is a high-fidelity PCC mathematical model of a pressure-vacuum swing adsorption process. We demonstrate that the cost-optimal design has limited process flexibility, challenging reactiveness to disturbances, such as those in the flue gas feed conditions. The results illustrate that flexibility can be introduced by relaxing the CO$_2$ recovery constraint on the operation, albeit at the expense of the capture efficiency of the process. We discover that adsorption-based processes can accommodate for significant flexibility and improved performance with respect to the operational constraints on CO$_2$ recovery and purity. The results herein demonstrate a trade-off between process economics and process operability, which must be effectively rationalised to integrate CO$_2$ capture units in the design of low-carbon energy systems.}

\maketitle
\doublespacing

\begin{figure}[htbp]
    \centering
    \includegraphics[width=1\textwidth]{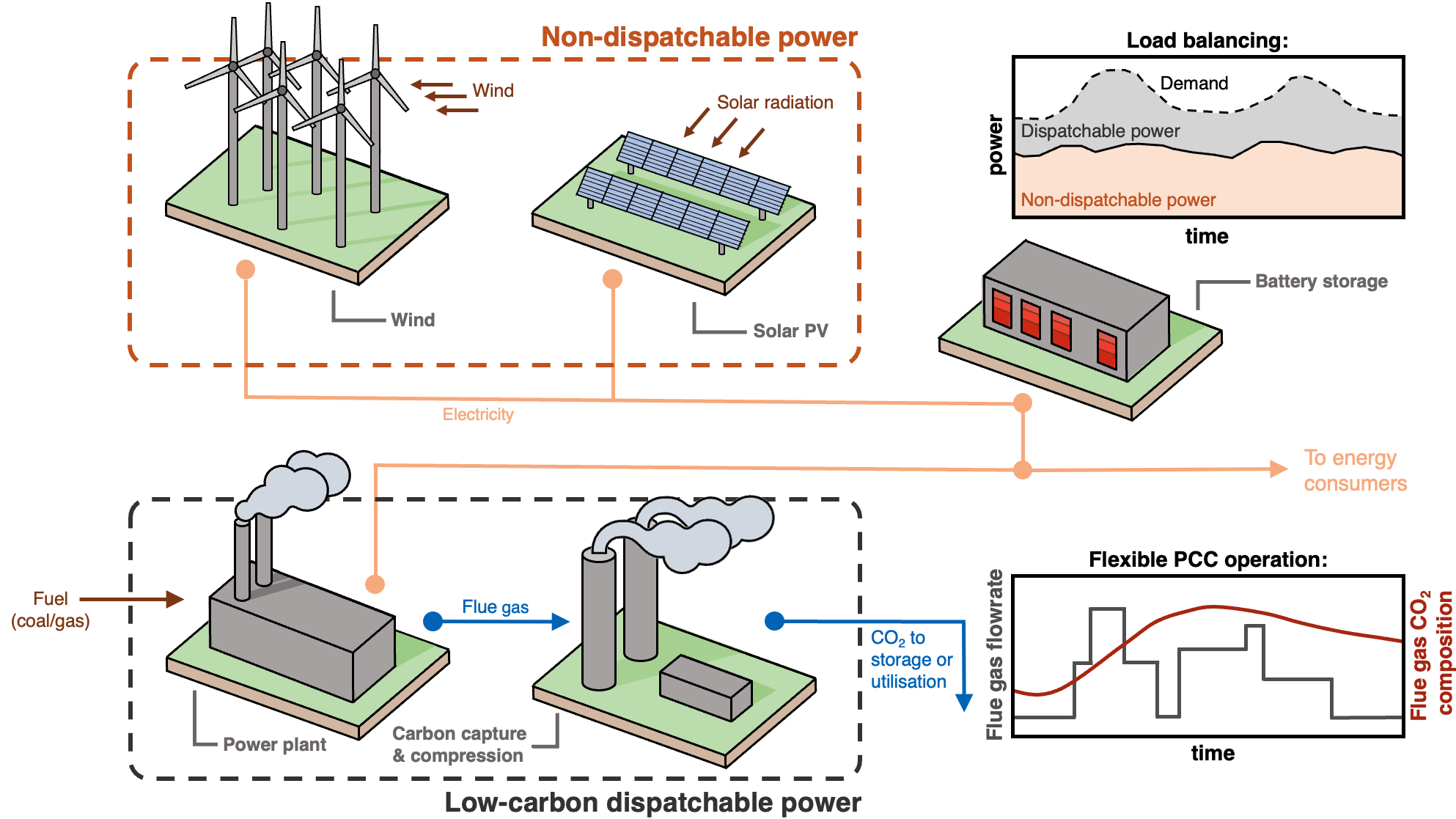}
    \caption{\textbf{Low-carbon energy system}. A schematic representation of a flexible low-carbon energy system with intermittent renewables (non-dispatchable power) and low-carbon dispatchable power. Power generation by fossil fuels is expected to provide a variable power output to match supply and demand of energy at any given time. Downstream post-combustion capture (PCC) is subject to operation under variable flue gas feed flowrate and CO$_2$ composition.}
    \label{fig:Conceptual_figure}
\end{figure}

\clearpage
As the world transitions to a low-carbon energy system, deployment of non-dispatachable renewable power, such as a wind and solar photovoltaic (PV), continues to expand \cite{IEA2021}. In Fig. \ref{fig:Conceptual_figure}, we illustrate a simplified low-carbon energy system in which baseline generation is provided by intermittent renewables, and load-balancing is handled by flexible low-carbon fossil fuel-fired power generation \cite{IPCC2022}. Such energy systems are crucial for the supply of continuous energy to enable uninterrupted industrial and residential use \cite{Pimm2019,Grasham2019,Diangelakis2016,Diangelakis2017}. A leading technology for dispatachable low-carbon power is combustion of fossil fuels with subsequent post-combustion carbon capture (PCC) and storage. In PCC and storage, CO$_2$ is removed from the resulting flue gas to yield a clean flue gas, which can be vented to the atmosphere, and a high purity CO$_2$ product stream \cite{Chao2021}. The latter can be compressed and sent to downstream processes for utilisation as a chemical feedstock \cite{Zhang2021, Jones2017}, or to be permanently sequestered in subsurface geological formations \cite{Krevor2023, Newell2019}. The highly dynamic nature of power plants operating in a load-balancing mode introduces variability in the flue gas stream in terms of both composition and flow rate \cite{Cristea2020, Patron2022, Mechleri2014}. Therefore, PCC processes need to be designed with a high degree of flexibility in mind to be robust under such conditions. This allows for fossil fuel power generation with PCC to become a reliable source of low-carbon dispatachable power in the emerging energy system \cite{Riboldi2017}.

The most mature technology for PCC is reversible chemical absorption into aqueous amine solutions \cite{Wang2011}. These processes are able to produce a high purity CO$_2$ product ($\geq95\%$) at high CO$_2$ recoveries (90-99\%). However, there are several practical challenges associated with the operation of these processes which lead to economic and environmental concerns, such as the large thermal energy requirement for regenerating the rich amine solution \cite{Meng2022}, thermal degradation of the solvent at the process operating conditions \cite{Rochelle2012}, and solvent losses owing to volatility \cite{Nguyen2010}. As a result of these challenges, research efforts have been dedicated towards exploring alternative technologies for carrying out the separation.

Adsorption-based processes represent an attractive alternative to amine-based absorption for PCC, primarily owing to the comparatively low energy penalty associated with regeneration of the sorbent \cite{Raganati2021}. In an adsorption-based process, a column is packed with pellets of a solid adsorbent material which is highly selective for concentrating CO$_2$ at its surface. An adsorption column is operated by cyclically varying the operating conditions to capture high purity CO$_2$ on the solid surface, and subsequently release it in a controlled manner \cite{Ruthven1984}. The most widely studied adsorption cycle for PCC is pressure-vacuum swing adsorption (PVSA) \cite{Bui2018, Webley2014, Nguyen2023, Khurana2019}, where adsorption is carried out at high pressure and the CO$_2$ product is extracted from the bed using a vacuum.  The design of PVSA carbon capture processes is typically conducted to minimise the cost per tonne of CO$_2$ captured, while satisfying constraints on the effectiveness of the separation \cite{Subraveti2021, Subraveti2022, Zanco2021, Peh2023}. To be suitable for geological storage, the CO$_2$ product stream must have a purity of at least 95\% \cite{Balashankar2019b, Burns2020}. It is generally an accepted standard, originally proposed by the United States Department of Energy (DoE), that post-combustion capture processes should attain a CO$_2$ recovery of at least 90\% \cite{Maruyama2020}. This approach for the design, according to minimisation of the capture cost, does not consider any aspects related to the flexibility of the process.

There is a clear need to move beyond the academic status-quo of designing for static, idealised scenarios, and incorporate flexibility into the design to ensure the long term viability of these processes throughout the energy transition. Firstly, there is currently very little available literature on the flexible operation of PVSA processes applied to PCC \cite{Wilkes2022}. In addition, we do not currently understand well how these processes respond to transient operating scenarios. Third, with analogy to other chemical processes, we expect there to be a trade-off between flexibility and process economics \cite{Pistikopoulos2020, Svensson2014, Tian2020}. Rationalising this trade-off is key to enabling industrial adoption of adsorption-based carbon capture processes.

In this work, we propose an approach for embedding flexibility assessment within the design of adsorption-based PCC systems to achieve low cost, while sufficiently prioritising operability from an early stage in the design. The approach utilises a high fidelity mathematical process model for the separation of CO$_2$/N$_2$ by PVSA and an associated techno-economic assessment to quantify the performance of PCC from a typical coal-fired power plant \cite{Ward2022b}. We have coupled the mathematical model with a framework for identifying the design space for which the process can be operated while satisfying regulatory requirements on the CO$_2$ product stream \cite{Sachio2022}. For any chosen operating strategy, the flexibility of the process can be quantified with respect to the design space boundary. In the following, we calculate and compare the flexibility resulting from several proposed PVSA design approaches. We find that the proposed direct design space approach is effective, efficient and provides a rich set of information around the process flexibility which is not given through classical process design.


\section*{Results}


\textbf{Case Study: Coal Fired Power Plant With CO$_2$ Capture}. In this work, we have analysed a four-step PVSA process, with feed pressurisation, applied for PCC from a typical coal fired power plant. The flue gas eluted from the power plant is considered to be a dry binary mixture of 15\%CO$_2$/85\%N$_2$ at a pressure of 1 bar and a temperature of 298 K. We consider an adsorption process utilising a packed bed of zeolite 13X adsorbent, a commercially available and widely studied material for PCC. The considered power plant has a typical specification, with a gross electrical power output of 1,000 MW and producing approximately 9 Mtn/yr of CO$_2$ emissions to the atmosphere. Full details of the process modelling and economic assessment conducted for this case study can be found in the methods Section.

\textbf{Process Behaviour in the Knowledge Space}. The problem is formulated as follows. In the first step, the design decisions (DDs) for the system are identified. These are the design parameters and/or operational variables which will be varied to understand the flexibility of the process. Here, we consider a typical PVSA design problem, where we use the feed velocity ($v_\mathrm{F}$), the high pressure ($p_\mathrm{H}$), and the intermediate pressure ($p_\mathrm{I}$) as DDs. In practice, implementation of PVSA processes on a large scale involves the operation of multiple columns in parallel. The columns are scheduled in such a way that the feed flue gas can be processed continuously. By-passing of flue gas and venting without capturing is not contemplated in this study. Therefore, in this setting, dynamically varying the cycling times ($t_\mathrm{ads}$ - adsorption time, $t_\mathrm{bd}$ - blowdown time, and $t_\mathrm{evac}$ - evacuation time) of the adsorption process to address disturbances is not realistic as this would result in significant process scheduling complications. Therefore, in this work, the cycle times are fixed. Upon varying the DDs of the system, we monitor the purity, recovery, energy consumption, productivity, and capture cost as key performance indicators (KPIs). In the second step, the bounds on the DDs are identified and used to generate the knowledge space (KSp). The KSp defines the sub-space of the entire design decision space for which we perform quasi-random sampling to probe the behaviour of the process. We have used the Sobol sequence to sample the design decision space and generate the KSp by taking 4,096 quasi-random samples. Of these samples, 3,458 satisfied the feasibility constraint on the operating pressures ($p_\mathrm{H}>p_\mathrm{I}$). For each of the sampled points, we record the DDs and respective KPIs of the process generated by running the process model and economic assessment.

\begin{figure}[htbp]
    \centering
    \includegraphics[width=0.7\textwidth]{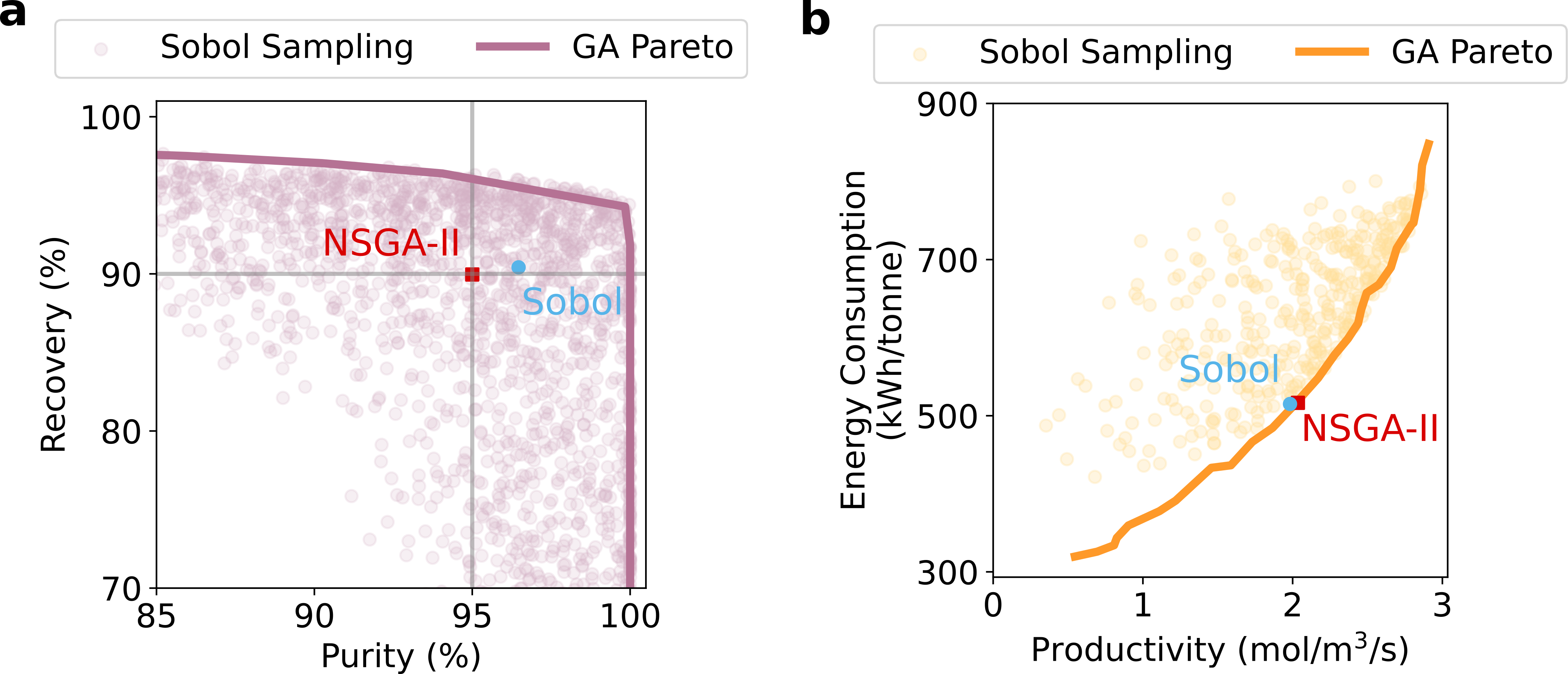}
    \caption{\textbf{Pareto front comparison between formal optimization (NSGA-II) and Sobol sampling.} \textbf{a} Unconstrained purity-recovery Pareto front. \textbf{b} Constrained energy-productivity Pareto front. In both \textbf{a} and \textbf{b}, the solid lines correspond to the NSGA-II Pareto fronts, while the scattered points correspond to Sobol sampling. The cost-optimal design of optimization using NSGA-II is highlighted as a red square in each Pareto plane, and that of the Sobol sampling is a blue circle. The corresponding design decisions and KPI values are summarised in Table \ref{Table:CostOptimal}.}
    \label{fig:Pareto}
\end{figure}

\textbf{Approximated Cost-Optimal Process Design}. Before performing design space identification and assessing the flexibility of the process, the KSp data set is used to approximate Pareto optimal frontiers of the process performance, and to approximate the cost-optimal operating point given the process performance constraints. As a benchmark of optimality, we have deployed the widely used non-dominated sorting genetic algorithm II (NSGA-II) to conduct rigorous process optimization. Using the NSGA-II routine, we have calculated Pareto fronts of unconstrained purity/recovery and constrained productivity/energy usage, as well as the constrained cost-optimal point. For the constrained problems, we require that the process produces CO$_2$ with purity $\geq$ 95\% and recovery $\geq$ 90\%, in compliance with regulatory requirements. In Fig. \ref{fig:Pareto}, we provide a comparison between the optimal process behaviour identified using NSGA-II (solid lines), and that obtained by sampling the KSp using the quasi-random Sobol sequence (symbols). The labelled points in each panel show the position in each Pareto plane of the cost-optimal point obtained by each method (blue: Sobol sampling, red: NSGA-II). We can see that there is excellent agreement between the Pareto fronts generated by each approach, validating the use of quasi-random sampling for the purposes of initial identification of the optimal process performance. In Table \ref{Table:CostOptimal}, we present the values of the DDs and minimum capture cost associated with the cost-optimal point identified using each approach. Again, we can see that the solutions are very similar, with the Sobol sampling approach obtaining a solution with a minimum capture cost deviating from the optimum by only 1.1\%. This is particularly impressive when considering the computational cost of each approach. In total, 15,120 forward simulations are performed for solving all three optimization problems to obtain the Pareto fronts and cost-optimal design (Fig. \ref{fig:Pareto}a, b) using NSGA-II. In comparison, only 3,458 forward simulations were required for Sobol sampling. We therefore contend that the Sobol sampling approach is an efficient and effective means to approximate the optimal performance of the adsorption process. Further to this, the outputs of Sobol sampling can further be used to generate a rich set of data around the flexibility of the process operation, as will be demonstrated below.

\begin{table}[htbp]
  \centering
  \caption{Design decisions and minimum capture cost for the cost-optimal solution from NSGA-II and Sobol sampling.}
  \label{Table:CostOptimal}
  \begin{tabularx}{0.95\textwidth}{
    >{\raggedright\arraybackslash}X
    >{\centering\arraybackslash}X
    >{\centering\arraybackslash}X
  }
  \hline
   & NSGA-II & Sobol sampling \\
  \hline
  \(\mathrm{p_{H}}\) (bar)       & 3.56  & 3.55  \\
  \(\mathrm{p_{I}}\) (bar)       & 1.72  & 1.25  \\
  \(\mathrm{v_{F}}\) (m/s)       & 1.44  & 1.40  \\
  Capture cost (\$/tonne)        & 61.67 & 62.35 (+1.1\%)*\\
  Purity (\%)                    & 95.0  & 96.5  (+1.6\%)*\\
  Recovery (\%)                  & 90.0  & 90.4  (+0.4\%)*\\
  Productivity (mol/m$^3$/s)     & 2.03  & 1.98  ($-$2.5\%)*\\
  Energy Consumption (kWh/tonne) & 516   & 515   ($-$0.2\%)*\\
  \hline
  \end{tabularx}
\begin{tablenotes}
   \item[*] The percentage change of key performance indicator with respect to the NSGA-II solution.  
  \end{tablenotes}
\end{table}

\begin{figure}[htbp]
    \centering
    \includegraphics[width=0.8\textwidth]{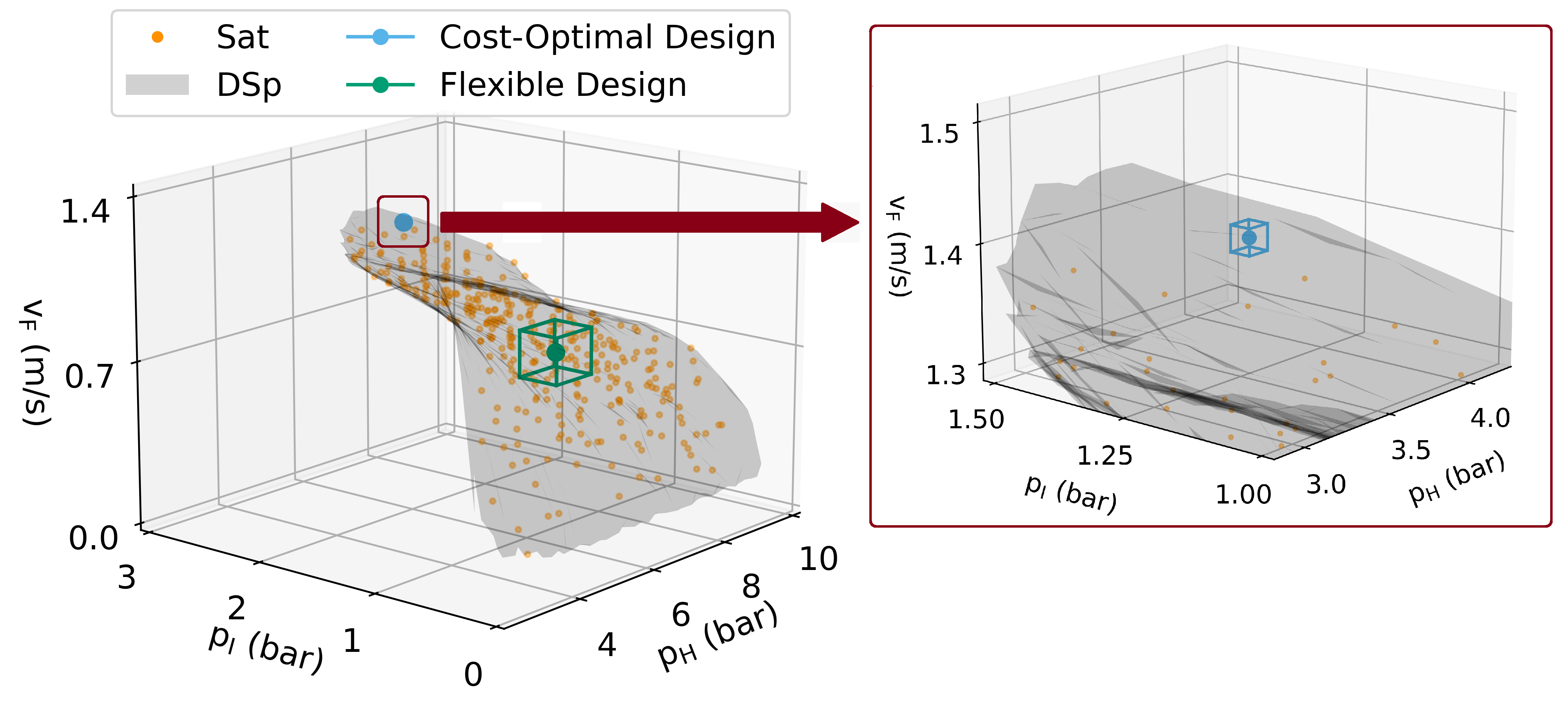}
    \caption{\textbf{Identification of the design space (DSp) and quantification of the acceptable operating region (AOR)}. The design space (DSp) representing the region of the KSp for which combinations of the design decisions satisfy CO$_2$ purity $\geq$ 95\% and recovery $\geq$ 90\% is shown as the shaded grey region. Points in the quasi-random sample which satisfy the constraints are shown in orange (Sat). Points which do not satisfy the constraints have been excluded for clarity. The nominal operating point (NOP) and corresponding acceptable operating region (AOR) for the cost-optimal design (blue) and the maximum flexibility design (green) are provided. The circle corresponds to the NOP and the box corresponds to the AOR.}
    \label{fig:Design_space_cost_flex}
\end{figure}

\textbf{Quantification of Acceptable Operating Ranges}. Following the design space identification framework established in \cite{Sachio2022}, an artificial neural network (ANN) is trained and used as an interpolator to increase the resolution of the KSp data set (see methods Section). Based on this high-resolution data set, we identify the design space via alpha shapes - as illustrated in Fig \ref{fig:Design_space_cost_flex}. The shaded region (grey) in Fig. \ref{fig:Design_space_cost_flex} is the design space whereby operation within it is guaranteed to satisfy the purity/recovery constraints. The mathematical definition of the design space enables quantitative analysis of flexibility with respect to any nominal operating point (NOP). Starting from an NOP of interest, a cuboid is formed and expanded outwards until one of its vertices hits the design space boundary. This forms an acceptable operating region (AOR) centered around the NOP. The side lengths of the AOR correspond to the multivariate proven acceptable range (MPAR) for each design decision, for which operation is guaranteed to be within the design space. Here, we investigate the cost-optimal design obtained above as the NOP of interest and identify its AOR, as shown in Fig. \ref{fig:Design_space_cost_flex}. We can see that the cost-optimal design has very low flexibility, as the AOR is not visible with reasonable axis scaling in Fig. \ref{fig:Design_space_cost_flex}. As it can be anticipated from the position of the cost-optimal design in the purity/recovery plane (Fig. \ref{fig:Pareto} (a)), there are active constraints at the cost-optimum point which cause this point to lie very close to the edge of the design space. This means that, in practice, the cost-optimal point would present low operability in the presence of disturbances to the nominal operation. The acceptable operating range is very limited, and the smallest disturbance in feed conditions ($\pm0.11\%$ with respect to nominal values of the DDs) would result in violation of operational constraints.

Through application of the design space identification framework, we are able to quantitatively ascertain that the status-quo design approach for PVSA processes yields an inflexible design with low operability in practice. At this stage, we seek to understand if there is scope in the design workflow to accommodate flexibility. More specifically, we aim at understanding how much flexibility can be allowed for in the design, and what impacts the accommodation of flexibility has on the other design outcomes - such as process efficiency and economics. To analyse this in detail, we will study two further design cases using the design space identification framework and the existing KSp data set. First, we will consider the ``relaxed cost-optimal design" - a case which maintains the original cost-optimal point, but allows for some relaxation of the recovery constraint to handle disturbances. Second, we will consider the ``maximum flexibility design" - a case which maximises flexibility within the original design space. In the following, we analyse each of these cases in further detail, and subsequently compare their performance.

\textbf{Flexibility by Constraint Relaxation}. In the design of PCC processes, the CO$_2$ product stream which is extracted from the flue gas must be of sufficient purity ($\geq 95\%$) to be suitable for geological sequestration. This design constraint is non-negotiable and cannot be violated, even for the purpose of improving process flexibility. However, the recovery constraint is only a target, and can potentially be violated if there are other operational benefits associated with doing so. Here, we study the case where we relax the recovery constraint to investigate the trade-off between flexibility and performance. The low discrepancy Sobol sampling coupled with the ANN surrogate model allows for the rapid assessment and identification of design spaces with different performance constraints. No additional high-fidelity model simulations are needed to characterize a new design space under different performance constraints. This showcases the usefulness of the design space identification framework for efficiently exploring different design options to attain process flexibility.

\begin{figure}[htbp]
    \centering
    \includegraphics[width=1\textwidth]{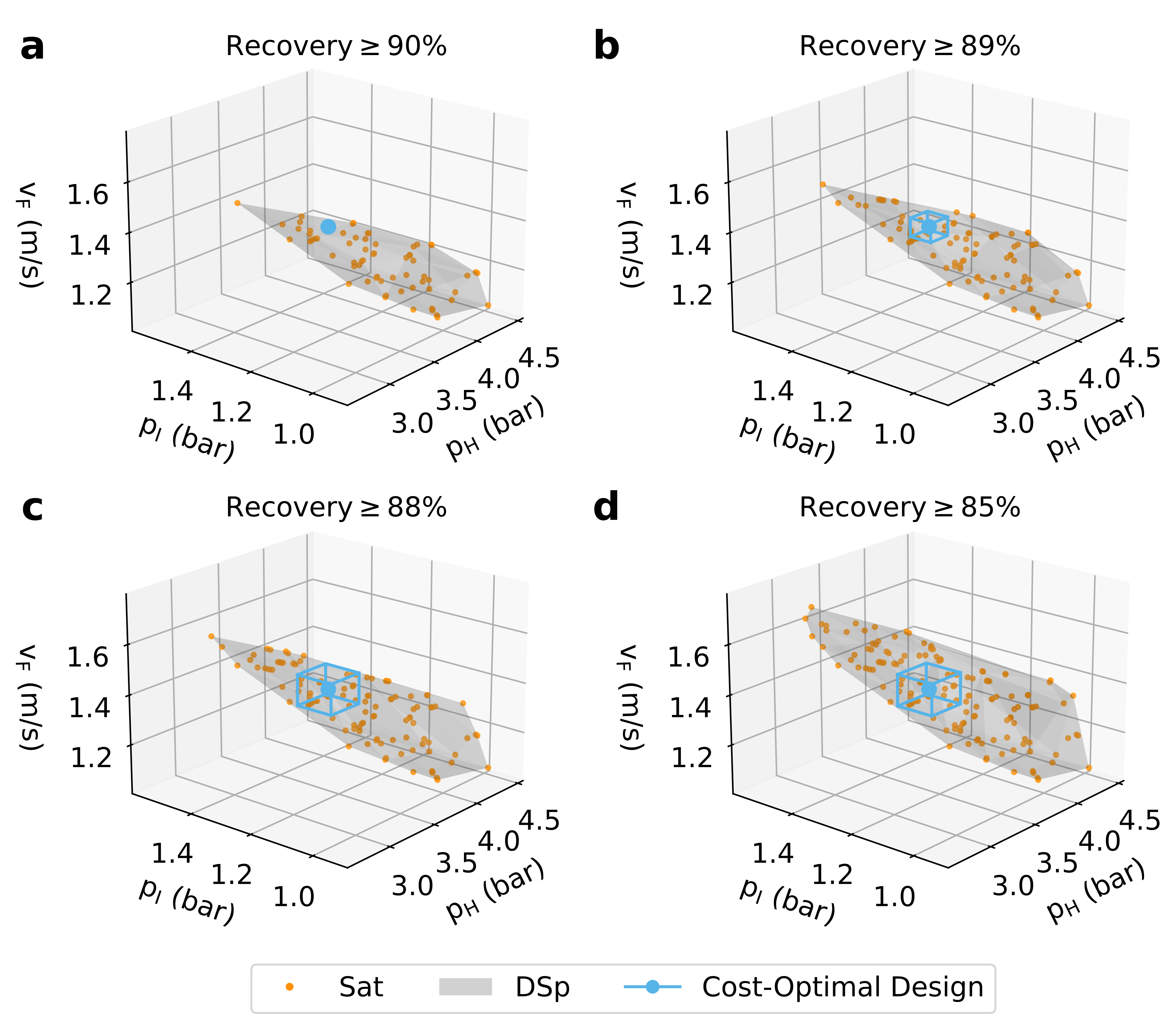}
    \caption{\textbf{Design spaces identification with progressive relaxation on the recovery constraint}. \textbf{a} Design space with the original recovery constraint ($\ge$ 90\%). \textbf{b} - \textbf{d} Design space with relaxed recovery constraint of 89\%, 88\%, and 85\%, respectively. The design space for each case is shown as the shaded grey region. The nominal operating point and corresponding acceptable operating region for each case are shown in blue. The quasi-random sampled points satisfying the constraints of each case are shown. The colour of each point corresponds to the capture cost of each operating point.}
    \label{fig:Relax_DSp}
\end{figure}

Fig. \ref{fig:Relax_DSp} shows the design spaces identified with different recovery targets, ranging from 85-90\%. As we can see, the design space expands predominately in the direction of increasing $p_\mathrm{H}$ as the recovery constraint is relaxed. When relaxing the recovery constraint to 89\% (1\% decrease from nominal target), the size of the AOR formed increased by 4 orders of magnitude (Fig. \ref{fig:Relax_DSp} (a) and (b)). The relaxation of the recovery constraint from 89\% to 88\% yields a further increase of the AOR (Fig. \ref{fig:Relax_DSp} (b) and (c)), while the latter does not change upon relaxing the constraint down to 85\% (Fig. \ref{fig:Relax_DSp} (c) and (d)). This behaviour can be explained by the change of the active constraint from recovery to purity. Recall, from Figure \ref{fig:Pareto} (a), that the obtained cost-optimal point from the Sobol sequence lies relatively further away from the purity constraint compared to the recovery constraint. This indicates that the recovery constraint is active, and is further proven by the increase in AOR size when the recovery constraint is relaxed. When the active constraint switches from recovery to purity, the region cannot expand further without violating the purity constraint because the identified AOR is centered around the cost-optimal design. It is noteworthy that if the cost-optimal solution from the NSGA-II were to be investigated as the NOP, both the recovery and purity constraints would need to be relaxed to see an increase in flexibility, as both constraints are active at that particular solution.

\begin{figure}[htbp]
    \centering
    \includegraphics[width=1\textwidth]{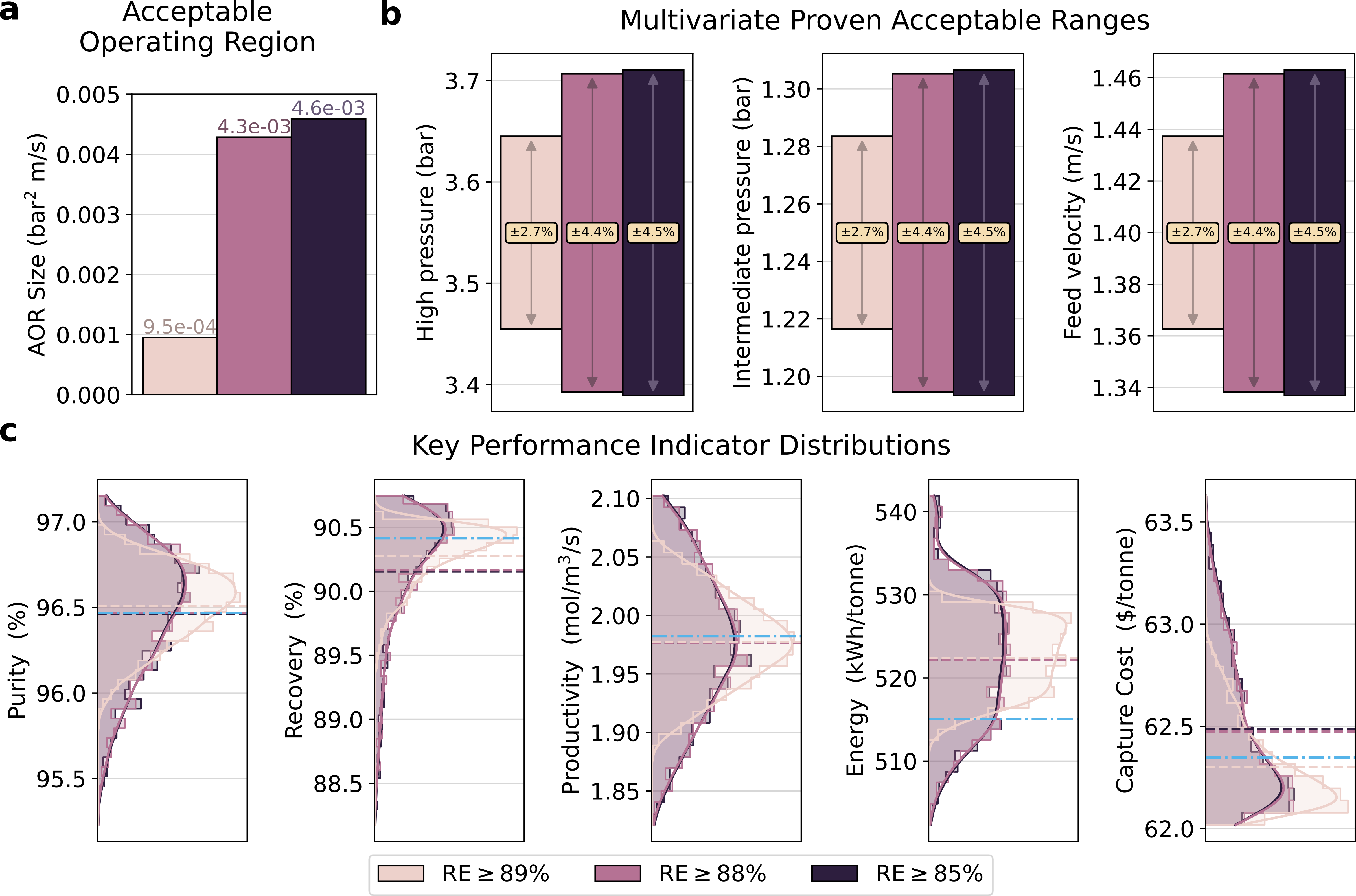}
    \caption{\textbf{Flexibility metrics comparison between three relaxed cost-optimal design cases} \textbf{a} Acceptable operating region (AOR) volume. \textbf{b} Multivariate proven acceptable ranges (MPARs) of the design decisions. \textbf{c} Distribution of all monitored KPIs within the identified acceptable operating region. The dashed line represents the mean value of the KPIs, while the blue dash dot line shows the KPI value at the cost-optimal point.}
    \label{fig:Relax_metrics}
\end{figure}

In Fig. \ref{fig:Relax_metrics}, the three relaxed cost-optimal designs are compared quantitatively. As anticipated above, the AOR size increases substantially ($\times$4.5) when the recovery constraint is relaxed from 89\% to 88\%, whereas the relative increase is quite modest (7\%) with a further relaxation of the constraint down to 85\% (Fig. \ref{fig:Relax_metrics} (a)). The corresponding multivariate proven acceptable ranges of the different cases are summarised in Fig. \ref{fig:Relax_metrics} (b). Generally, all three cases have relatively low flexibility, whereby a maximum disturbance of $\leq 4.5\%$ on the three operating variables ($v_\mathrm{F}$, $p_\mathrm{H}$ and $p_\mathrm{I}$) can be accommodated by the design. In Fig. \ref{fig:Relax_metrics} (c), we show the corresponding distributions of the process KPIs within the AOR. We find that the KPI distributions for the 88\% and 85\% cases are essentially overlapping, in accordance with these two cases having virtually identical AORs. Not surprisingly, the mean of the distribution of recovery is reduced with increasing relaxation of the recovery constraint, and the distribution becomes more skewed towards lower recovery values. On the contrary, the distribution of purity expands more symmetrically and its mean remains the same for each design (including the original cost-optimal design). This observation supports our previous hypothesis that purity becomes the active constraint in the relaxed designs. The behaviour of the mean of the distribution of the energy usage and capture cost (both increasing with relaxation) appears less intuitive, as one would expect the relaxed scenarios to consume less energy. However, these scenarios carry higher \emph{specific} energy demand and cost, because less CO$_2$ is captured. Yet, the 88\% recovery case has a mean capture cost of 62.47 \$/tonne which is only 2\% higher than the cost-optimal solution, while offering an AOR that is 63\% larger than that of the 89\% recovery case. Hence, this could represent an effective method for increasing process flexibility.

The framework has successfully quantified the impact of performance constraint relaxation on the process flexibility. In this case, a relaxation of the recovery constraint to 85\% is not considered to be a worthy trade-off in the design, given the diminishing return on process flexibility. It is also worth noting that while this strategy does allow for a moderate increase in the process flexibility, it also demands that there is some decrease in the amount of CO$_2$ emissions captured from the power plant flue gas. Ultimately, at the industrial-scale, such violation may not be desirable from an environmental perspective, as small changes in the recovery can lead to significant increases in the absolute plant emissions.


\textbf{Flexibility by Design}. We have used an iterative quasi-random grid search approach to find the design which gives the largest possible AOR, while meeting the original purity/recovery constraints. As shown in Fig.  \ref{fig:Design_space_cost_flex}, we can see that this design that maximises flexibility lies at the center of the design space. The capture cost of the flexible design is 70.16 \$/tonne, which represents an increase of 12.5\% relative to the cost-optimal design. Therefore, we can infer that there is an inherent trade-off in the design between capture cost and flexibility. This trade-off will need to be effectively rationalised in the design workflow to yield carbon capture processes which are both economical and operable. Notably, different from the relaxed design case, it will be shown in the following that the higher cost of this new design case is associated with an improved performance of the separation. 

\begin{figure}[htbp]
    \centering
    \includegraphics[width=1\textwidth]{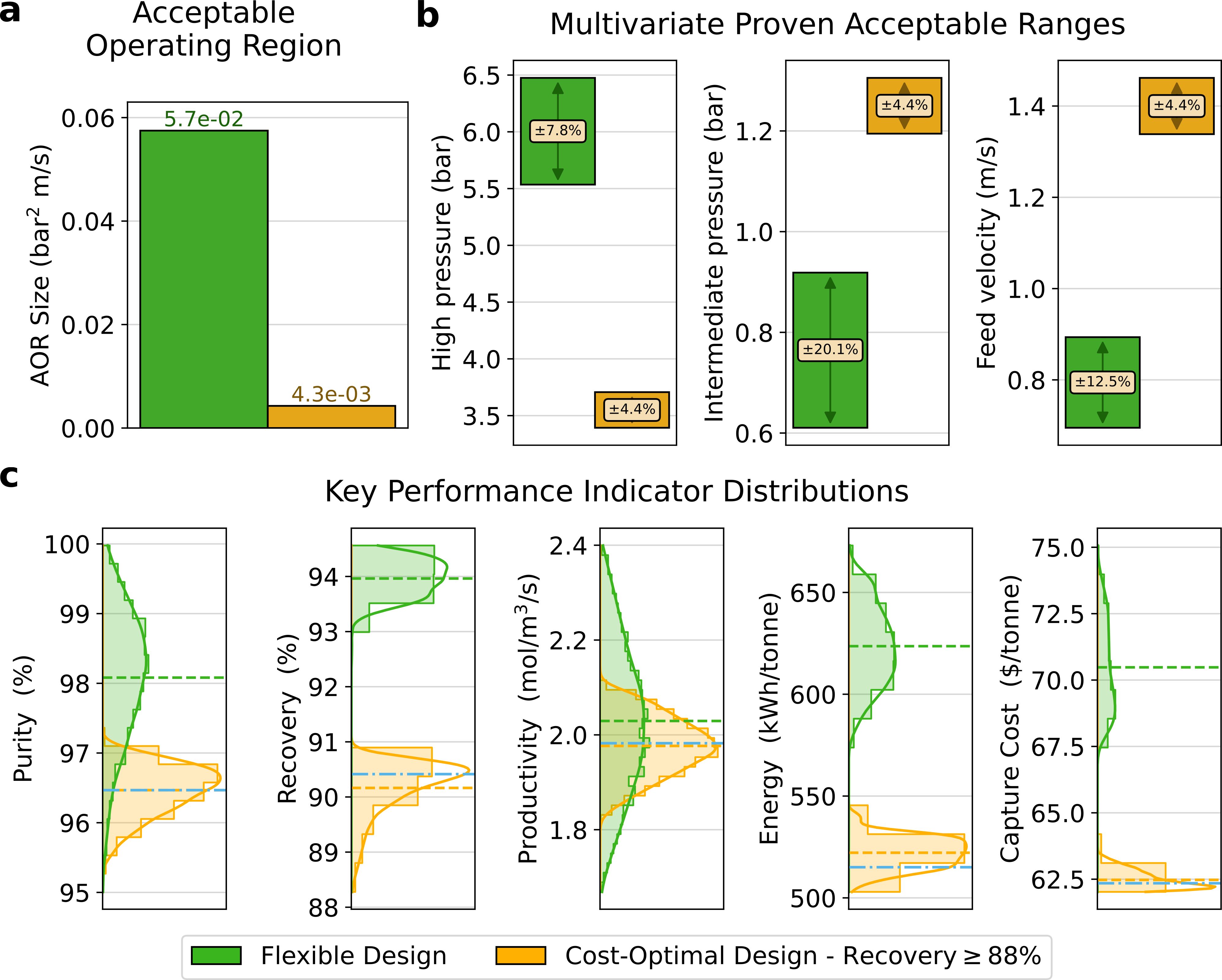}
    \caption{\textbf{Flexibility metrics comparison between the maximum flexibility design and the relaxed cost-optimal design} \textbf{a} Acceptable operating region (AOR) size. \textbf{b} Multivariate proven acceptable ranges (MPARs) of the design decisions. \textbf{c} Distribution of all monitored KPIs within the identified AOR. The dashed line represents the mean value of the KPIs, while the blue dash dot line shows the KPI value at the cost-optimal point.}
    \label{fig:DesCompKPI}
\end{figure}

We compare in Fig. \ref{fig:DesCompKPI} the most flexible design and the relaxed cost optimal design (recovery $\geq 88\%$). The most flexible design offers an AOR that is 13-fold larger than that of the relaxed cost-optimal design (Fig. \ref{fig:DesCompKPI} (a)). The larger AOR translates directly onto the width of the MPARs for $p_\mathrm{H}$, $p_\mathrm{I}$ and $v_\mathrm{F}$, as shown in Fig. \ref{fig:DesCompKPI} (b). The most flexible design can now accommodate larger variations in the design decisions, namely of up to 10\% ($p_\mathrm{H}$ and $v_\mathrm{F}$) and 20\% ($p_\mathrm{I}$). The cost-operability trade-off is visible also in the relative location of each design decision. For example, relative to the relaxed cost optimum design, the flexible design has a higher high pressure ($p_\mathrm{H}$), and a lower intermediate pressure ($p_\mathrm{I}$). The lower value of $p_\mathrm{I}$ generates higher purity for the process by rejecting more nitrogen from the bed during blowdown. The higher value of $p_\mathrm{H}$ increases the recovery of CO$_2$ by increasing the CO$_2$ affinity during the adsorption step. However, increasing the value of $p_\mathrm{H}$ is achieved by operating the feed compressor at a higher pressure, increasing the electricity usage of the process. We thus expect the capture cost of the flexible design to be larger than the relaxed cost optimal design.

The distribution of the KPIs for these two design cases offers additional insight into the cost-operability trade-off (Fig. \ref{fig:DesCompKPI} (c)). Across all KPIs, the flexible design is shown to have larger distribution ranges. This is expected as a larger AOR encompasses more combinations of the design decisions. For example, the nominal productivity of the two designs is similar (approx. 2 mol/m$^3$/s), but the most flexible design shows a standard deviation of 0.15 mol/m$^3$/s, which is 2.5x larger than the value observed for the cost-optimal designs. Notably, the purity and recovery distributions for the relaxed cost-optimal design push against the respective constraints, whereas the most flexible design displays purity and recovery distributions shifted towards enhanced separation performance. To generate additional flexibility for the most flexible process design, the grid search algorithm needs to move the NOP away from the boundary of the design space. This means that the purity and recovery achieved by the process need to exceed the strict requirements to give space around the NOP. This is a key co-benefit of the most flexible design, wherein the flexibility of the process comes with a higher specific capture cost (13\% higher) and specific energy consumption (20\% greater) - but the performance of the separation is also increased. The most flexible design prevents a greater amount of CO$_2$ emissions from the power plant from reaching the atmosphere and the outlet CO$_2$ stream is purer and could therefore be suitable for use in a wider variety of downstream utilisation processes.

\section*{Conclusion}
We have analysed the operational flexibility of an adsorption-based post-combustion CO$_2$ capture process through a design space identification framework. We discover that the design approach of minimising capture cost yields a process that is inherently inflexible. In practical scenarios of transient flue gas production, such design would fail to meet the commonly adopted purity-recovery (95\%–90\%) constraints on the produced CO$_2$. We propose and compare two alternative design approaches aimed at introducing operational flexibility. First, we consider relaxation of the recovery constraint (down to 85\%), while retaining the cost-optimal design as the nominal operating point. This approach yields moderate flexibility (variation in the operating variables of up to 4.5\%), but also demands that the capture rate of the plant decreases during disturbed operation (up to 7\%). Second, we identify the operating point with maximum flexibility in the design space and observe that the process can accommodate for variations in the operating variables of up to 10-20\%. However, the capture cost increases by 12.5\% relative to the cost-optimal design, because this added flexibility is achieved by exceeding the purity-recovery constraints. 

The transition to a sustainable and reliable energy system will encompass a growing proportion of baseline generation provided by intermittent renewables with load-balancing handled by low-carbon fossil fuel-fired power generation. As such, the latter must be associated with a CO$_2$ capture process that can accommodate for disturbances in flue gas conditions resulting from the load balancing requirements. The results herein demonstrate a trade-off between process economics and process operability, which must be effectively rationalised to integrate CO$_2$ capture units in low-carbon energy systems. In this context, future work must consider the integration of constraints related to the process operability \textit{within} the design framework to arrive at capture processes which both have an acceptable capture cost, and are sufficiently flexible to handle highly transient flue gas production. Such information should be obtained by detailed model-based control studies using realistic process disturbance data as an input to understand the required level of flexibility to arrive at a controllable process design and operational scheme.


\section*{Methods}


\subsection*{PVSA process model}

The high-fidelity mathematical model used to describe the PVSA process analysed in this study was developed during previous studies \cite{Ward2022a, Ward2022b}, where full details of the implementation can be found. The model has been extensively validated against both experimental dynamic column breakthrough data available in the literature \cite{Pini2021, Casas2012, Jee2001}, and independent dynamic adsorption simulations \cite{Pini2021, Haghpanah2013}. The model simulates the performance of a four-step pressure-vacuum swing adsorption cycle with feed pressurisation \cite{Haghpanah2013}. The steps of the cycle are 1) high-pressure adsorption, 2) forward blowdown, 3) reverse evacuation and 4) feed pressurisation. To model the adsorption column dynamics, we have described the flow of an ideal gas mixture through a packed bed of adsorbent pellets using the axially dispersed plug flow model. This is coupled with a solid-phase adsorption kinetics model which expresses the rate of adsorption into the packed bed as a first order mass transfer process through the linear driving force approximation, assuming that the rate of adsorption is limited by the diffusion of adsorbate molecules in the inter-crystalline macro-pores of the adsorbent \cite{Haynes1973} and adopting the modification of Hassan et al. \cite{Hassan1985} to allow for extension of the expression into the non-linear region of the adsorption isotherm. The pressure drop in the column has been described using Darcy's law. The energy balance equations describe several relevant mechanisms of heat transfer in both the gas-phase and the column wall, including conduction, convection, heat released through exothermic adsorption and heat losses to the environment. The adsorption equilibrium of CO$_2$/N$_2$ on zeolite 13X has been described using the extended dual-site Langmuir isotherm model \cite{Haghpanah2013}. The governing balance equations of mass, momentum and energy are given in non-dimensional form as follows. The overall mass balance is:

\begin{equation}
    \frac{\partial \bar{p}}{\partial \tau} - \frac{\bar{p}}{\bar{T}}\frac{\partial \bar{T}}{\partial \tau} = -\bar{T}\frac{\partial}{\partial Z} \left( \frac{\bar{p}\bar{v}}{\bar{T}} \right) - \psi\bar{T}\sum_{i=1}^{n_\mathrm{c}}\frac{\partial x_i}{\partial \tau}
\end{equation}

The component mass balance is:

\begin{equation}
    \frac{\partial y_i}{\partial \tau} + \frac{y_i}{\bar{p}}\frac{\partial \bar{p}}{\partial \tau} - \frac{y_i}{\bar{T}}\frac{\partial \bar{T}}{\partial \tau} = \frac{1}{\mathrm{Pe}}\frac{\bar{T}}{\bar{p}}\frac{\partial}{\partial Z}\left( \frac{\bar{p}}{\bar{T}} \frac{\partial y_i}{\partial Z} \right) - \frac{\bar{T}}{\bar{p}}\frac{\partial}{\partial Z} \left( \frac{y_i\bar{p}\bar{v}}{\bar{T}} \right) - \frac{\bar{T}}{\bar{p}}\psi\frac{\partial x_i}{\partial \tau}
\end{equation}

The rate of adsorption is:

\begin{equation}
    \frac{\partial x_i}{\partial \tau} = \alpha_i (x_i^* - x_i)
\end{equation}

The pressure drop is:

\begin{equation}
    -\frac{\partial \bar{p}}{\partial Z} = \frac{150}{4r_\mathrm{p}^2}\left( \frac{1-\epsilon}{\epsilon} \right)^2 \frac{v_0 L}{p_0} \mu \bar{v}
\end{equation}

The gas-phase energy balance is:

\begin{equation}
    \frac{\partial \bar{T}}{\partial \tau} + \Omega_2 \frac{\partial \bar{p}}{\partial \tau} = \Omega_1\frac{\partial^2 \bar{T}}{\partial Z^2} - \Omega_2 \frac{\partial}{\partial Z}\left( \bar{p}\bar{v} \right) + \sum_{i=1}^{n_\mathrm{c}} \left[ \left( \sigma_i - \Omega_3\bar{T} \right) \frac{\partial x_i}{\partial \tau} \right] - \Omega_4 \left( \bar{T} - \bar{T}_\mathrm{w} \right)
\end{equation}

And the column wall energy balance is:

\begin{equation}
    \frac{\partial \bar{T}_\mathrm{w}}{\partial \tau} = \Pi_1 \frac{\partial^2 \bar{T}_\mathrm{w}}{\partial Z^2} + \Pi_2 \left( \bar{T} - \bar{T}_\mathrm{w} \right) - \Pi_3 \left( \bar{T}_\mathrm{w} - \bar{T}_\mathrm{a} \right)
\end{equation}

The governing balance equations are a coupled set of partial differential equations (PDEs) which describe variation in the column state variables as a function of both space and time. To solve the balance equations, the PDEs have been discretised with respect to space using the weighted essentially non-oscillatory (WENO) finite volume scheme to yield a system of coupled time-dependant ordinary differential equations (ODEs) \cite{Haghpanah2013, Medi2011, Cruz2005}. This set of ODEs has been solved in MATLAB using the \textit{ode15s} solver. The integration of the equations is carried out subject to boundary conditions representing the four-step adsorption cycle \cite{Haghpanah2013, Ward2022b}. Due to the cyclic nature of the adsorption process, the balance equations must be integrated several times until a cyclic steady state (CSS) is attained, defined by a material balance error of less than 0.5\% over the previous five cycles for both CO$_2$ and N$_2$. We therefore define CSS using \cite{Wilkins2022}:

\begin{equation}
    100\%\times \bigg\vert \frac{\left( n_i^\mathrm{in} - n_i^\mathrm{out} \right)}{n_i^\mathrm{in}} \bigg\vert \leq 0.5 \%, \quad \forall i
\end{equation}

Once CSS has been attained, the trajectories of the state variables as a function of space and time across a complete cycle can be used to calculate the performance of the adsorption process through the evaluation of key performance indicators (KPIs). We have used the purity $(\mathrm{Pu}_\mathrm{CO_2})$ and recovery $(\mathrm{Re}_\mathrm{CO_2})$ of the extracted CO$_2$, the process productivity $(\mathrm{Pr})$, the specific energy usage $(E_\mathrm{T})$, and the capture cost $(C_\mathrm{CO_2}^\mathrm{cap})$ as KPIs, which are calculated as:

\begin{equation}\label{purity}
    \mathrm{Pu}_\mathrm{CO_2}\,(\%) = 100\times \frac{n_{\mathrm{CO}_2\mathrm{,out}}^{\mathrm{evac}}}{n_{\mathrm{CO}_2\mathrm{,out}}^{\mathrm{evac}} + n_{\mathrm{N}_2\mathrm{,out}}^{\mathrm{evac}}}
\end{equation}

\begin{equation}\label{recovery}
    \mathrm{Re}_\mathrm{CO_2}\,(\%) = 100\times \frac{n_{\mathrm{CO}_2\mathrm{,out}}^{\mathrm{evac}}}{n_{\mathrm{CO}_2\mathrm{,in}}^{\mathrm{pres}} + n_{\mathrm{CO}_2\mathrm{,in}}^{\mathrm{ads}}}
\end{equation}

\begin{equation}
    \mathrm{Pr} \, (\mathrm{mol/m^3/s}) = \frac{n_{\mathrm{CO}_2\mathrm{,out}}^{\mathrm{evac}}}{V_\mathrm{bed} \cdot t_\mathrm{cycle}} 
\end{equation}

\begin{equation}
    E_\mathrm{T} \, (\mathrm{kWh/tonne}) = \frac{E_\mathrm{ads} + E_\mathrm{bd} + E_\mathrm{evac} + E_\mathrm{pres}}{m_{\mathrm{CO}_2\mathrm{,out}}^{\mathrm{evac}}}
\end{equation}

\begin{equation}
    C_\mathrm{CO_2}^\mathrm{cap} \, (\mathrm{\$/tonne}) = \frac{\mathrm{Total\,annual\,cost (\$/yr)}}{\mathrm{Re_{CO_2}} \cdot \dot{m}_\mathrm{CO_2}^\mathrm{emitted} \mathrm{(tonne/yr)}}
\end{equation}


\subsection*{Economic assessment}

The PVSA process model has been coupled to a detailed economic assessment to evaluate the cost per tonne of CO$_2$ captured for post-combustion CO$_2$ capture from the flue gas of a typical 1,000 MW coal fired power plant. Details of the modelling parameters used as inputs to define this case study can be found in the Supporting Information file. The economic assessment model used in this study was developed as part of a previous study \cite{Ward2022b}. Full details of the case study and the associated economic calculations can be found in the previous work. Below, we briefly outline the three stages of the economic assessment model. First, the adsorption process is scaled up from the operation of a single adsorption column, as simulated by the process model, to a full-scale set of parallel columns which is able to continuously accept the full volume of flue gas produced by the power plant, by scheduling individual adsorption columns into a number of parallel trains \cite{Subraveti2021}. Second, the capital cost the process equipment is estimated from correlations available in the literature \cite{Turton2018}. We consider the major equipment costs, which are those of the adsorption columns, compressors and vacuum pumps, to calculate the total direct cost (TDC) of the equipment \cite{Effendy2017}. We use a bottom-up approach to convert the TDC to the total capital required (TCR), which additionally accounts for costs associated with developing the project and allowing sufficient contingencies \cite{Subraveti2021}. We then apply the capital recovery factor method to calculate the equivalent annual cost (EAC), which can be used in conjunction with the capture rate to calculate the cost of capital per tonne of CO$_2$ captured \cite{Zanco2021}. Finally, we calculate the operating cost of the process through the contributions of electricity usage by the compressors and vacuum pumps, as well as the cost of annual adsorbent replenishment to account for continuous sorbent degradation \cite{Effendy2017, Severino2021}. We combine the annualised capital cost and the operating cost together to calculate the overall cost per tonne of CO$_2$ captured.


\subsection*{Process optimisation}

When conducting formal optimisation of the performance of the PVSA process, we have deployed the non-dominated sorting genetic algorithm II (NSGA-II). We have utilised an implementation of the NSGA-II routine which is available as the \textit{ga} function in the MATLAB Global Optimisation Toolbox. When executing the NSGA-II routine, we have used a population size of 72 and we run the algorithm for a maximum of 70 generations. The design decisions and associated parametric bounds used during optimization of the process performance can be found in the Supporting Information file.


\subsection*{Design space identification}

The design space identification framework deployed in this work was originally developed in a previous work \cite{Sachio2022}. The framework comprises of three main steps: 1. problem formulation, 2. design space identification, and 3. flexibility assessment. In step 1, a process of interest is characterized using mathematical models that can be of any form (e.g., mechanistic, data-driven). Then, the design decisions that are of interest are defined. Depending on the objective, these can include design parameters, operation variables, and process disturbances. Next, are the monitored key performance indicators (KPIs), feasibility constraints, and performance constraints.

In step 2, the quasi-random Sobol sampling is performed as per the problem definition to generate the knowledge space. From this data set, a neural network is trained and used to increase the resolution of the data set. Then, the constraints are imposed to characterize the satisfied points and violated points. A bisection-based algorithm is then used to calculate an alpha shape with an alpha radius that characterizes the design space without violations inside of it. Finally, in step 3, the identified design space is utilized to quantify acceptable operating regions with respect to any nominal operating point. This is also done using a bisection-based algorithm where we expand a cuboid region from the nominal operating point of interest until one of the vertices hits the edge of the design space. Based on this, the multivariate proven acceptable ranges can be extracted, and distributions of the monitored KPIs can be evaluated. We have developed an open-source Python package which can be used for steps 2 \& 3 of the framework (\emph{'dside'}: https://github.com/stvsach/dside).


\subsection*{Quasi-random sampling for Knowledge Space Generation}
In this work, the knowledge space generation is performed based on the quasi-random Sobol sequence \cite{Sobol1993}. The Sobol sequence is used to generate the combinations of design decisions in Python using \textit{scipy.stats.qmc.Sobol} class from the \textit{scipy} library. This can also be done using the SobolGSA software developed by Kucherenko et al \cite{Kucherenko2013,Kucherenko2021}. The sequence is a quasi-random technique that aims to reduce discrepancy within the sampled space. This enables the training of an artificial neural network that comprehends the knowledge space as a whole and is capable to perform satisfactory interpolation. The sequence is used to generate 4,096 design decision combinations, which are then scaled appropriately with respect to the design decision bounds of the study. Then, the feasibility constraint tied to the high and intermediate operating pressures ($p_\mathrm{H}\geq p_\mathrm{I}$) is imposed to filter the infeasible input combinations. A total of 3,458 input combinations remain after the screening and are used to run the high-fidelity model simulations in a parallel fashion. When sampling the knowledge space, the following parametric sampling bounds are applied; $p_\mathrm{H} \in [1, 10]$ bar, $p_\mathrm{I} \in [0.05, 5]$ bar and $v_\mathrm{F} \in [0.1, 2]$ m/s. We fix the remaining operating parameters of the system at the following values; $p_\mathrm{L} = 0.03$ bar, $t_\mathrm{ads} = 50$ s and $t_\mathrm{bd} = t_\mathrm{evac} = 100$ s.


\subsection*{Data-driven resolution support}
The design space is defined as a region within the knowledge space where all constraints are satisfied. The design space identification framework relies on the calculation of alpha shapes, which are geometrical hulls \cite{Edelsbrunner1994three} that can be used to define the design space boundary. In order to characterize design spaces with smooth boundaries and no internal violations, it is crucial to utilise a high resolution data set \cite{Sachio2022}. To acquire a higher resolution data set, one could perform more sampling based on the high-fidelity model. However, this would incur significant computational burden because the mathematical process model is computationally expensive and requires several minutes of CPU time to acquire a single data point. Therefore, we have deployed an artificial neural network (ANN) surrogate model to act as an interpolator on the data set generated by sampling the high-fidelity model for the purpose of increasing the resolution of the sampled data set to enable accurate characterisation of the design space. In this work, three separate neural networks are trained to predict the different KPIs, given the combination of design decisions ($\mathrm{p_H}$, $\mathrm{p_I}$, and $\mathrm{v_F}$). The first ANN predicts purity and recovery, the second predicts energy consumption and productivity, and finally, the third predicts the capture cost. The inputs and outputs for all of the ANNs are normalized using the min/max method. The KSp data set was shuffled and split 90\%-10\% into training/testing sets, respectively. Each neural network uses a feed-forward architecture, with 3 hidden layers and 256 hidden units. Each hidden unit applies the rectified linear unit (ReLU) function to its inputs. The network architecture was determined by a grid search over the possible network architectures. The weight/bias parameters for the networks have been trained using the adaptive moments (ADAM) optimisation algorithm with a learning rate of $\alpha = 5\times10^{-5}$ for 50,000 epochs. The training is performed in Python using the \textit{pytorch} library.

\section*{Acknowledgements}
We thank Prof Cleo Kontoravdi for their input on the design space framework. This work has received financial support from the Industrial Decarbonisation Research and Innovation Centre (IDRIC). S.S. and A.W. are grateful for funding by departmental scholarships from the Department of Chemical Engineering, Imperial College London. A.W. also acknowledges financial support by the Burkett Scholarship.

\section*{Author Contributions}
M.M.P. and R.P. conceptualized and supervised the project. A.W. and S.S. designed the methodology. A.W. carried out the numerical simulations of the adsorption process. S.S. carried out design space analysis. A.W. and S.S. wrote the original draft of the manuscript. M.M.P. and R.P. reviewed and edited the manuscript.   

\section*{Data and Code Availability}
The source data for all figures are available from the corresponding author upon reasonable request. The code used for the design space identification framework is available from GitHub (https://github.com/stvsach/dside).


\newpage
\singlespacing
\bibliography{sn-bibliography}

\end{document}